\documentclass[12pt]{article}
\usepackage{graphicx} 

\usepackage{url}

\usepackage[square,numbers]{natbib}
\usepackage{hyperref}
\usepackage{xcolor}

\usepackage{orcidlink}

\usepackage{hhline}
\usepackage{tikz}
\usepackage{graphicx}
\usepackage{pgfplots}
\usetikzlibrary{shapes,positioning,calc,arrows,automata,intersections,matrix,decorations,math}
\usetikzlibrary{arrows.meta}

\definecolor{bblue}{HTML}{4F81BD}
\definecolor{rred}{HTML}{C0504D}
\definecolor{ggreen}{HTML}{9BBB59}
\definecolor{ppurple}{HTML}{9F4C7C}

\definecolor{red1}{HTML}{FF4B00}
\definecolor{blue1}{HTML}{005AFF}
\definecolor{green1}{HTML}{03AF7A}
\definecolor{sky1}{HTML}{4DC4FF}
\definecolor{orange1}{HTML}{F6AA00}
\definecolor{yellow1}{HTML}{FFF100}

\usepackage{pifont}

\usepackage{tabu}
\usepackage{here}
\usepackage{tabularx}
\usepackage{multirow}
\usepackage{booktabs}
\usepackage{array}

\usepackage{bbm}
\usepackage{amsfonts}
\usepackage{mathtools}
\usepackage{bm}
\usepackage{url}

\usepackage{amsmath}
\usepackage{amsthm}

\usepackage{authblk}

\DeclareMathOperator*{\argmin}{arg\,min}

\newcommand{\Rset}{\mathbb{R}}

\newcommand{\boldSigma}{\boldsymbol{\Sigma}}

\newcommand{\transpose}{^{\mathrm{T}}}
\newcommand{\diag}{\mathop{\text{diag}}}

\newcommand{\diagM}{\mathop{\text{diagMat}}}

\newcommand{\mat}[1]{\textbf{#1}}

\title{Implicit ZCA Whitening Effects of Linear Autoencoders for Recommendation}
\author[1]{Katsuhiko Hayashi\ \orcidlink{0000-0002-3240-4697}}
\author[2]{Kazuma Onishi}
\affil[1]{Faculty of IST, Hokkaido University}
\affil[2]{School of Engineering, Hokkaido University}

\date{}

\begin{document}

\maketitle

\begin{abstract}
Recently, in the field of recommendation systems, linear regression~(autoencoder) models have been investigated as a way to learn item similarity.
In this paper, we show a connection between a linear autoencoder model and ZCA whitening for recommendation data.
In particular, we show that the dual form solution of a linear autoencoder model actually has ZCA whitening effects on feature vectors of items, while items are considered as input features in the primal problem of the autoencoder/regression model.
We also show the correctness of applying a linear autoencoder to low-dimensional item vectors obtained using embedding methods such as Item2vec to estimate item-item similarities.
Our experiments provide preliminary results indicating the effectiveness of whitening low-dimensional item embeddings.
\end{abstract}

\section{Introduction}
E-commerce has become an indispensable part of our daily lives. Recommender systems help users find products they want to buy on e-commerce sites and have a wide range of applications, such as in movie recommendations and product recommendations. Collaborative filtering~(CF) is one of the most widely used approaches in recommender systems. 
Nearest neighbor CF approaches are divided into user-based and item-based ones.
In item-based collaborative filtering~(ICF), recommendations to users are made by finding items that are similar to other items that a given user has already had an interaction with. Therefore, the similarity measure between two items plays an important role in the ICF approach.
While early ICF models used such statistical measures as Pearson correlation and cosine similarity~\cite{deep,advances,partial},
model-based methods~\cite{slim,fism,ease} have recently been investigated as a way to learn item similarity. 
In particular, ICF models based on linear regressions~(autoencoders)~\cite{slim,ease,towards,diag} have achieved current state-of-the-art performances on several benchmark datasets for implicit recommendation.

As to the reason for the empirical success of linear autoencoders, in this study, 
we reveal that they actually have ZCA whitening~\cite{whiten} effects on recommendation data.
The whitening transformation removes correlations between the feature dimensions of the item vectors, and the item-item similarity estimated from the transformed vectors improves
the quality of the recommendations.
Our finding also has the following empirical contribution:
\begin{itemize}
\item {\bf Whitening Item Embeddings}: In the field of natural language processing, it is known that whitening word embeddings improves the performance of various retrieval and similarity tasks~\cite{whiteningbert1,whiteningbert2}. Analogously, embedding methods such as Item2vec~\cite{item2vec}, which learns a latent semantic feature vector representation of items, are useful in the field of information retrieval.
Thus, whitening item embeddings is expected to improve recommendation performance. Our experiments provide preliminary results that show the effectiveness of whitening low-dimensional item embeddings.
\end{itemize}

\paragraph{Notation and Preliminaries}
Vectors are represented by
boldface lowercase letters, e.g., \mat{a}.
$\mat{0}_D$ and $\mat{1}_D$ are $D$-dimensional vectors of zeros and ones, respectively.
Matrices are represented by boldface capital letters, e.g., \mat{A}.
The $i$-th row of a matrix \mat{A} is represented by \mat{a}$_{i:}$, and the
$j$-th column of \mat{A} is represented by \mat{a}$_{:j}$.
The element $(i,j)$ of a matrix \mat{A} is denoted by $a_{ij}$.
$\mat{A}\transpose$ and $\mat{A}^{-1}$ denote the transpose and inverse of a matrix \mat{A}, respectively.
$\mat{I}_D$ denotes the $D$-dimensional identity matrix.
$\diag(\mat{A})$ is the diagonal of a square matrix $\mat{A}$.
$\diagM(\mat{a})$ denotes the diagonal matrix whose diagonal is the vector $\mat{a}$.


\section{Item-based Neighborhood Model}
Let $U$ and $I$ be sets of users and items, respectively.
Like in many papers on recommender systems~\cite{slim,ease,partial}, we consider implicit feedback data.
Here, the user-item interaction matrix $\mat{X}$ can be considered to be a binary one:
\[
\mat{X}=\left( \begin{array}{cccc}
x_{11} & x_{12} & \cdots & x_{1|I|}\\
x_{21} & x_{22} & \cdots & x_{2|I|} \\
\vdots & \vdots & \ddots & \vdots \\
x_{|U|1} & x_{|U|2} & \cdots & x_{|U||I|}
\end{array} \right) \in \{0,1\}^{|U|\times |I|}.
\]
where $x_{ui} = 1$ represents that there is an interaction
between user $u$ and item $i$. If there is no interaction between $u$ and $i$, then $ x_{ui} = 0$.

To make recommendations for a user~$u$, item-based neighborhood collaborative filtering~(ICF) models~\cite{advances}
require pre-computed similarities associated with each item-item pair from $\mat{X}$.
We denote an item-item similarity matrix as $\mat{B}\in\Rset^{|I|\times|I|}$, where $b_{ij}$ is the similarity between two items $i$ and $j$.
The item set that the user $u$ has interacted with is represented by $\mat{y}_{(u)}\in\{0,1\}^{|I|}$, where the $j$-th element is 1 if there is an interaction between $u$ and $j$, otherwise 0.
Accordingly, ICF models simply compute user $u$'s preference score $s_{ui}$ on item $i$ as the following dot-product~$s_{ui}=\mat{y}_{(u)}\transpose\mat{b}_{:i}$,
where $\mat{b}_{:i}$ is the $i$-th column of $\mat{B}$.

\section{Shallow Linear Autoencoders}
As the previous section shows, a key
step in ICF methods is to estimate
the item-item similarity matrix $\mat{B}\in\Rset^{|I|\times|I|}$ from $\mat{X}\in\{0,1\}^{|U|\times|I|}$.
While early ICF approaches used such statistical measures as Pearson correlation and cosine similarity~\cite{deep,advances,partial},
model-based methods~\cite{slim,fism,ease} have recently been investigated as a way to learn item similarity.
In this section, we introduce shallow linear autoencoder models~\cite{slim,ease} that learn the item-item similarity matrix as a regression problem.

\subsection{Linear Autoencoder with L2 Regularization}
Linear autoencoders can be trained with the multivariate least squares fitting approach.
The linear algebraic formulation can be represented as a ridge linear regression problem:
\begin{equation}
\label{eq:autoencoder}
\widehat{\mat{B}}=\argmin_{\mat{B}}\Bigl\{||\mat{X}-\mat{X}\mat{B}||_{F}^{2}+\lambda||\mat{B}||_{F}^{2}\Bigr\}
\end{equation}
where $\lambda>0$ is a regularization parameter.
The closed form solution of the above equation is given as
\begin{equation}
\label{eq:solution1}
\widehat{\mat{B}} = (\mat{X}\transpose\mat{X}+\lambda\mat{I}_{|I|})^{-1}\mat{X}\transpose\mat{X}
\end{equation}
or,  equivalently~\cite{dual},
\begin{equation}
\label{eq:solution2}
\widehat{\mat{B}} = \mat{X}\transpose(\mat{X}\mat{X}\transpose+\lambda\mat{I}_{|U|})^{-1}\mat{X}.
\end{equation}
For a proof of the equivalence between Eqs.(\ref{eq:solution1}) and (\ref{eq:solution2}), we refer the read to the Appendix A of the paper~\cite{iterative}~(in the case of $|I|>|U|$) and our Appendix~\ref{append:A} (in the case of $|U|>|I|$).
Note that though the minimization of Eq.(\ref{eq:autoencoder}) is achieved in an obvious way~($\mat{B}=\mat{I}_{|I|}$), $\diag{(\mat{B})}=\mat{0}_{|I|}$ is imposed as a constraint condition in practical linear autoencoder models like SLIM~\cite{slim} and EASE~\cite{ease}. 

\subsection{EASE: Linear Autoencoder with \\ Diagonal Constraints~\cite{ease}}
Steck~\cite{ease} introduced a linear autoencoder model, called EASE.
The objective function of EASE for learning the item-item similarity matrix $\mat{B}$ is: 
\begin{eqnarray}
\label{eq:ae}
&&\widehat{\mat{B}}_{\text{EASE}}=\argmin_{\mat{B}}\Bigl\{||\mat{X}-\mat{X}\mat{B}||_{F}^{2}+\lambda||\mat{B}||_{F}^{2}\Bigr\} \nonumber\\
&&\text{s.t.}\ \ \ \diag{(\mat{B})}=\mat{0}_{|I|}.\nonumber
\end{eqnarray}
The optimization problem can be solved with the following Lagrangian:
\[
L=||\mat{X}-\mat{X}\mat{B}||_{F}^{2}+\lambda||\mat{B}||_{F}^{2}+2\boldsymbol{\alpha}\transpose\diag{(\mat{B})}
\]
where $\boldsymbol{\alpha}=[\alpha_1,\dots,\alpha_{|I|}]\transpose$ is the vector of Lagrange multipliers.
By setting the derivative to zero,
we derive the estimate of the similarity matrix $\mat{B}$:
\begin{equation*}
\widehat{\mat{B}}_{\text{EASE}} = (\mat{X}\transpose\mat{X}+\lambda\mat{I}_{|I|})^{-1}(\mat{X}\transpose\mat{X}-\diagM{(\boldsymbol{\alpha})}).
\end{equation*}
By imposing the constraint $\diag(\widehat{\mat{B}}_{\text{EASE}}) = 0$ and defining $\widehat{\mat{P}} = (\mat{X}\transpose\mat{X}+\lambda\mat{I}_{|I|})^{-1}$, $\boldsymbol{\alpha}$ is determined as
\begin{equation*}
\boldsymbol{\alpha}=\mat{1}_{|I|}\oslash\diag(\widehat{\mat{P}})-\lambda\mat{1}_{|I|}.
\end{equation*}
Steck~\cite{ease} showed that the solution can be derived in the following closed form:
\begin{equation*}
\label{eq:solutionease}
\widehat{\mat{B}}_{\text{EASE}} = \mat{I}_{|I|}-\widehat{\mat{P}}\diagM{(\mat{1}_{|I|}\oslash\diag(\widehat{\mat{P}}))}
\end{equation*}
where $\oslash$ denotes elementwise division.

\section{Relationship between Linear Autoencoder \\ and ZCA Whitening}
\subsection{ZCA Whitening~(Zero-phase Component Analysis)}
Whitening is an operation that eliminates correlations between features in a data sample. That is, correlation between any two components $x_i$ and $x_j$ of a sample vector $\mat{x}=[x_1,\dots,x_D]$ in the $D$-dimensional feature space is reduced to zero. In the following, we assume that a feature vector of each sample is centered; i.e., $\frac{1}{D}\sum_{i=1}^{D}x_i=0$.
Given $N$ data samples~$\mat{X}=[\mat{x}_1,\dots,\mat{x}_N]\in\Rset^{D\times N}$,
the correlation coefficients between features are represented by the following covariance matrix:
\[
\mat{$\Phi$}_{\mat{X}}=\frac{1}{N}\mat{X}\mat{X}\transpose.
\]

We now consider a linear transformation $\mat{P}\in\Rset^{D\times D}$ of feature vectors:
\[
\mat{w}_n=\mat{P}\mat{x}_n\ \ \ (n=1,\dots,N).
\]
Accordingly, the covariance matrix of the transformed vectors $\mat{W}=[\mat{w}_1,\dots,\mat{w}_N]$ is defined as:
\[
\mat{$\Phi$}_{\mat{W}}=\frac{1}{N}\mat{W}\mat{W}\transpose.
\]
The purpose of whitening is to find a projection matrix $\mat{P}$ that makes the covariance matrix $\mat{$\Phi$}_{\mat{W}}$ be the identity matrix $\mat{I}_D$.
If $\mat{$\Phi$}_{\mat{W}}=\mat{I}_D$, $\mat{P}$ must satisfy the following equation:
\begin{equation}
\label{eq:whitencond}
\mat{P}\transpose\mat{P}=\mat{$\Phi$}_{\mat{X}}^{-1}.
\end{equation}
$\mat{P}$ that satisfies Eq.(\ref{eq:whitencond}) can be represented by using the eigenvectors $\mat{U}$ of $\mat{X}\mat{X}\transpose$.
\begin{equation*}
\mat{X}\mat{X}\transpose=\mat{U}\boldSigma\mat{U}\transpose
\end{equation*}
where $\mat{U}$ is a square $D\times D$ matrix whose $i$-th column is the $i$-th eigen vector of $\mat{X}\mat{X}\transpose$, and $\boldSigma$ is a diagonal matrix whose diagonal elements are the corresponding eigenvalues, $d_{ii}=\lambda_i$.
Since $\boldSigma$ is an orthogonal matrix ($\mat{U}\transpose\mat{U}=\mat{U}\mat{U}\transpose=\mat{I}_D$), $\mat{$\Phi$}_{\mat{X}}^{-1}$ can be denoted by $\mat{U}\boldSigma^{-1}\mat{U}\transpose$.
In zero-phase component Analysis~(ZCA) or ZCA whitening~\cite{whiten}, we use 
\[
\mat{P}_{\text{ZCA}}=\mat{U}\boldSigma^{-1/2}\mat{U}\transpose
\]
as a transformation matrix $\mat{P}$.
In practice, noise $\epsilon$ is often added to the diagonal of $\boldSigma$:
\begin{equation*}
\mat{P}_{\text{ZCA}}=\mat{U}(\boldSigma+\epsilon\mat{I}_{D})^{-1/2}\mat{U}\transpose.
\end{equation*}

\subsection{Whitening Effects of Linear Autoencoders}
\label{sec:whiten-auto}
\paragraph{Linear Autoencoder with L2 Regularization}
First, we consider the ZCA whitening transformation of the user-item interaction matrix $\mat{X}\in\{0,1\}^{|U|\times|I|}$~(here, we assume $|U|>|I|$):
\[
\mat{W}=\mat{P}_{\text{ZCA}}\mat{X}
\]
where $\mat{P}_{\text{ZCA}}=\mat{U}(\boldSigma+\epsilon\mat{I}_{|U|})^{-1/2}\mat{U}\transpose$.
Note that $\mat{U}$ and $\boldSigma$ are the eigen vector and eigenvalue matrices of $\mat{X}\mat{X}\transpose$.
After this whitening transformation,
we simply consider
\[
\widehat{\mat{B}}_{\text{ZCA}}=\mat{W}\transpose\mat{W}
\]
as an item-item similarity matrix for ICF-based recommender systems.

Now let us rewrite $\widehat{\mat{B}}_{\text{ZCA}}$:
\begin{eqnarray*}
\widehat{\mat{B}}_{\text{ZCA}} &=& \mat{W}\transpose\mat{W} = (\mat{P}_{\text{ZCA}}\mat{X})\transpose(\mat{P}_{\text{ZCA}}\mat{X}) \\
&=& \mat{X}\transpose \bigl(\mat{U}(\boldSigma+\epsilon\mat{I}_{|U|})^{-1/2}\mat{U}\transpose\bigr) \bigl(\mat{U}(\boldSigma+\epsilon\mat{I}_{|U|})^{-1/2}\mat{U}\transpose\bigr) \mat{X} \\
&=& \mat{X}\transpose \bigl(\mat{U}(\boldSigma+\epsilon\mat{I}_{|U|})^{-1}\mat{U}\transpose\bigr)\mat{X} \\
&=& \mat{X}\transpose \bigl(\mat{U}\boldSigma\mat{U}\transpose + \epsilon\mat{U} \mat{U}\transpose\mat{U}\mat{U}\transpose\bigr)^{-1}\mat{X}\\
&=& \mat{X}\transpose \bigl(\mat{X}\mat{X}\transpose+\epsilon\mat{I}_{|U|}\bigr)^{-1}\mat{X}\ \ \ \ \ \ (=\text{Eq.(\ref{eq:solution2})}) \\
&=& \bigl(\mat{X}\transpose\mat{X}+\epsilon\mat{I}_{|I|}\bigr)^{-1}\mat{X}\transpose\mat{X}\ \ \ \ \ \ (=\text{Eq.(\ref{eq:solution1})}).
\end{eqnarray*}
Refer to our Appendix~\ref{append:A} for details on the final transformation.
This result clearly shows a connection between the shallow linear autoencoder and ZCA whitening.
We can see that linear autoencoders actually have implicit ZCA whitening-like effects on feature vectors of items\footnote{In general, linear autoencoder models for recommendation data do not assume that input data are centered.}, while
items are considered as input features in the primal problem of the autoencoder models.

\paragraph{Linear Autoencoder with Diagonal Constraints}
As pointed out in the paper~\cite{diag}, the solution of EASE can be divided into two terms: regularization and diagonal constraints. The former regularization part is equivalent to the
solution of linear autoencoder with L2 regularization:
\[
\widehat{\mat{B}}_{\text{EASE}}=\underbrace{\bigl(\mat{X}\transpose\mat{X}+\lambda\mat{I}_{|I|}\bigr)^{-1}\mat{X}\transpose\mat{X}}_{\widehat{\mat{B}}_{\text{ZCA}}}-\bigl(\mat{X}\transpose\mat{X}+\lambda\mat{I}_{|I|}\bigr)^{-1}\diagM(\boldsymbol{\alpha}).
\]
This result shows that EASE also has implicit whitening effects on item vectors.
From \cite{diag}, it is known that the latter diagonal constraint part plays a role that penalizes the impact of unpopular items.
In future work, we will further investigate the role of the diagonal constraint part from the view point of the data preprocessing stage.

\paragraph{Whitening Item Embeddings}
Several studies~\cite{item2vec,wmf,towards} have shown that a latent semantic representation of items is useful for estimating item similarity.
We consider item embeddings $\mat{E}\in\Rset^{D\times|I|}$, where $D<|I|$ is the dimension size of the item embedding.
Our finding on a relationship between the linear autoencoder and ZCA whitening shows the correctness of using a linear autoencoder on item embeddings:
\[
\widehat{\mat{B}}=\argmin_{\mat{B}}\Bigl\{||\mat{E}-\mat{E}\mat{B}||_{F}^{2}+\lambda||\mat{B}||_{F}^{2}\Bigr\}.
\]
The final solution is
\[
\widehat{\mat{B}} = (\mat{E}\transpose\mat{E}+\lambda\mat{I}_{|I|})^{-1}\mat{E}\transpose\mat{E}.
\]
In the case of $D<|I|$, by using Lemma 9 of the paper~\cite{iterative}, we can transform the above equation into
\[
\widehat{\mat{B}} = \mat{E}\transpose(\mat{E}\mat{E}\transpose+\lambda\mat{I}_{D})^{-1}\mat{E}.
\]
Moreover, we can further transform this equation as follows:
\begin{eqnarray*}
\widehat{\mat{B}} &=& \mat{E}\transpose(\mat{E}\mat{E}\transpose+\lambda\mat{I}_{D})^{-1}\mat{E} \\
&=& \mat{E}\transpose(\mat{U}\boldSigma\mat{U}\transpose+\lambda\mat{U}\mat{I}_{D}\mat{U}\transpose)^{-1}\mat{E} \\
&=& \mat{E}\transpose\mat{U}(\boldSigma+\lambda\mat{I}_{D})^{-1}\mat{U}\transpose\mat{E} \\
&=& \mat{E}\transpose\mat{U}(\boldSigma+\lambda\mat{I}_{D})^{-1/2}\mat{U}\transpose\mat{U}(\boldSigma+\lambda\mat{I}_{D})^{-1/2}\mat{U}\transpose\mat{E} \\
&=& (\mat{P}_{\text{ZCA}}\mat{E})\transpose(\mat{P}_{\text{ZCA}}\mat{E})
\end{eqnarray*}
where $\mat{U}\boldSigma\mat{U}\transpose$ is the eigenvalue decomposition of $\mat{E}\mat{E}\transpose$ and $\mat{P}_{\text{ZCA}}$ denotes $\mat{U}(\boldSigma+\lambda\mat{I}_D)^{-1/2}\mat{U}\transpose$.
The result shows that the linear autoencoder implicitly decorrelates latent features of item embeddings through the ZCA whitening process.

\section{Experiments}
\subsection{Datasets and Evaluation Metrics}
We conducted experiments on two publicly available datasets:
\begin{itemize}
\item MovieLens 20 Million (ML-20M)~\cite{ml}: 136,677 users
and 20,108 movies with about 10.0 million interactions,
\item Netflix Prize (Netflix)~\cite{netflix}: 463,435 users and 17,769 movies with about 56.9 million interactions,
\end{itemize}
For a fair comparison, we followed the experimental settings used in~\cite{vae} and kept the same pre-processing
steps\footnote{The program code is provided by the authors of~\cite{ract}: \url{https://github.com/samlobel/RaCT_CF/blob/master/setup_data.py}.}.

The experiments considered two ranking metrics, Recall$@R$ and the truncated NDCG (NDCG$@R$), where $R$
is a cut-off hyper-parameter~\cite{vae}.
Here, $\omega(r)$ is defined as the item at rank $r$, $\mathbbm{I}[\cdot]$ as the indicator function, and $\mathcal{I}_u$ as the held-out unobserved items that a user $u$ will interact.
Recall$@R$ for user $u$ is
\[
\text{Recall}@R:=\sum_{r=1}^{R}\frac{\mathbbm{I}[\omega(r)\in \mathcal{I}_u]}{\min(R,|\mathcal{I}_u|)}.
\]
The truncated discounted cumulative gain (DCG$@R$) is
\[
\text{DCG}@R:=\sum_{r=1}^{R}\frac{2^{\mathbbm{I}[\omega(r)\in \mathcal{I}_u]}-1}{\log{(r+1)}}.
\]
NDCG$@R$ is obtained by dividing DCG$@R$ by its best possible value where all the held-out items are ranked at the top.

\subsection{Results}
\begin{table}[t]
\centering
\small
\caption{Comparison of recommendation performances: * and ** denote results transcribed from~\cite{vae} and~\cite{ease}, respectively.}
\label{tab:results}
\begin{tabular}{rrrrr}
\toprule
      & \multicolumn{2}{c}{\textbf{ML-20M}} & \multicolumn{2}{c}{\textbf{Netflix}} \\
Model & Recall@20 & NDCG@100 & Recall@20 & NDCG@100
\\\midrule
{\bf Embed}~(Inner Product) & 0.279 & 0.317 & 0.246 & 0.279 \\
{\bf Embed+AE} & 0.372 & 0.402 & 0.353 & 0.385 \\
{\bf Embed+EASE} & 0.361 & 0.394 & 0.341 & 0.375 \\\midrule
*SLIM~\cite{slim} & 0.370 & 0.401 & 0.347 & 0.379 \\
**EASE~\cite{ease} & 0.391 & 0.420 & {\bf 0.362} & {\bf 0.393} \\
*WMF~\cite{wmf} & 0.360 & 0.386 & 0.316 & 0.351 \\
*CDAE~\cite{denoise} & 0.391 & 0.418 & 0.343 & 0.376 \\
*Mult-VAE~\cite{vae} & 0.395 & 0.426 & 0.351 & 0.386 \\
RaCT~\cite{ract} & {\bf 0.403} & {\bf 0.434} & 0.357 & 0.392 \\
\bottomrule
\end{tabular}
\ \\
\end{table}
In section~\ref{sec:whiten-auto}, we showed that linear autoencoders have ZCA whitening-like effects on item feature vectors.
In our experiments, we investigated how linear autoencoders improve the quality of item embeddings.
Using a singular value decomposition $\mat{X}=\mat{U}\boldSigma\mat{V}\transpose$, we computed item embeddings $\mat{E}=\boldSigma^{1/2}\mat{V}\transpose\in\Rset^{D\times|I|}$.
To estimate item similarity, we applied a linear autoencoder with L2 regularization~({\bf AE}) and {\bf EASE} to item embeddings, respectively.
We also tried a simple inner product $\mat{E}\transpose\mat{E}$ as a baseline item-item similarity matrix.

The number of dimensions $D$ was set to 800 and 4,000 for the ML-20M and
Netflix datasets, respectively. 
The hyperparameter $\lambda$ of the linear autoencoders was fixed to 200 in all settings.
Note that an advantage of using low-dimensional vectors is the computational savings regarding the calculation of the inverse matrix; i.e.,
$(\mat{E}\mat{E}\transpose+\mat{I}_{D})^{-1}\in\Rset^{D\times D}$ is much more efficient than
$(\mat{X}\transpose\mat{X}+\lambda\mat{I}_{|I|})^{-1}\in\Rset^{|I|\times|I|}$.

Tab.\ref{tab:results} compares recommendation performances.
Our results clearly show that linear autoencoders improve the quality of the item embeddings.
In particular, {\bf Embed+AE} outperformed {\bf Embed+EASE}.
We think this is because a low-rank approximation (item embeddings) eliminates the effects of unpopular items, while the diagonal constraint part in the solution of EASE also plays the same role, as noted in Section~\ref{sec:whiten-auto}.
Therefore, we think that the normal {\bf AE} already has sufficient ability to improve the quality of the item embeddings.

\section{Conclusion}
In this paper, we have shown that linear autoencoder models have ZCA whitening-like effects on recommendation data.
This finding ensures the correctness of applying a linear autoencoder model to low-dimensional item embedding vectors. Our initial experiments also reveal the effectiveness of whitening low-dimensional item embeddings.

In future work, we will try other methods to make item embeddings, such as Item2vec~\cite{item2vec} and GloVe~\cite{glove}.

\bibliographystyle{abbrvnat}
\bibliography{papers}

\appendix

\section{Proof of the Equivalence of Eqs.(\ref{eq:solution1}) and (\ref{eq:solution2}) when $|U|>|I|$}
\label{append:A}
\begin{proof}
\ \\
Given a matrix $\mat{X}\in\Rset^{|U|\times|I|}$, we assume $|U|>|I|$. 

Let $\mat{X}=\mat{U}_f\boldSigma_f\mat{V}^{T}$ be the full SVD representation of $\mat{X}$ with $\mat{U}_f\transpose\mat{U}_f=\mat{U}_f\mat{U}_f\transpose=\mat{I}_{|U|}$ and $\mat{V}\transpose\mat{V}=\mat{V}\mat{V}\transpose=\mat{I}_{|I|}$.
Further, $\boldSigma_f=\begin{pmatrix}
\boldSigma \\
\mat{0}
\end{pmatrix}\in\Rset^{|U|\times|I|}$ and $\mat{U}_f=(\mat{U}\ \mat{U}_{\perp})\in\Rset^{|U|\times|U|}$
where $\boldSigma\in\Rset^{|I|\times|I|}$ is a diagonal matrix whose diagonal entries are the singular values of $\mat{X}$ and $\mat{U}_{\perp}\in\Rset^{|U|\times(|U|-|I|)}$ 
denotes the matrix of the bottom $|U|-|I|$
left singular vectors.

We can rewrite:
\[
\mat{X}\mat{X}\transpose+\lambda\mat{I}_{|U|}=\mat{U}_f(\boldSigma_f\boldSigma_f\transpose+\lambda\mat{I}_{|U|})\mat{U}_f\transpose.
\]
Then, we have:
\begin{eqnarray}
\label{eq:a}
\widehat{\mat{B}} &=& \mat{X}\transpose(\mat{X}\mat{X}\transpose+\lambda\mat{I}_{|U|})^{-1}\mat{X} \nonumber\\
&=& \mat{X}\transpose(\mat{U}_f(\boldSigma_f\boldSigma_f\transpose+\lambda\mat{I}_{|U|})^{-1}\mat{U}_f\transpose)\mat{X} \nonumber\\
&=& \mat{V}\boldSigma_f\transpose(\boldSigma_f\boldSigma_f\transpose+\lambda\mat{I}_{|U|})^{-1}\mat{U}_f\transpose\mat{X}.
\end{eqnarray}
Here, we rewrite:
\begin{eqnarray}
\label{eq:b}
\boldSigma_f\transpose(\boldSigma_f\boldSigma_f\transpose+\lambda\mat{I}_{|U|})^{-1} &=& (\boldSigma\ \mat{0})\begin{pmatrix}
(\boldSigma^2+\lambda\mat{1}_{|I|})^{-1} & \mat{0} \\
\mat{0} & \frac{1}{\lambda}\mat{I}_{|U|-|I|}
\end{pmatrix} \nonumber\\
&=& (\boldSigma(\boldSigma^2+\lambda\mat{1}_{|I|})^{-1}\ \mat{0})\in\Rset^{|I|\times|U|}.
\end{eqnarray}
We combine Eqs.(\ref{eq:a}) and (\ref{eq:b}):
\begin{eqnarray*}
\label{eq:c}
\widehat{\mat{B}} &=& \mat{V}(\boldSigma(\boldSigma^2+\lambda\mat{1}_{|I|})^{-1}\ \mat{0})\mat{U}_f\transpose\mat{X} \\
&=&  \mat{V}(\boldSigma(\boldSigma^2+\lambda\mat{1}_{|I|})^{-1}\ \mat{0})(\mat{U}\ \mat{U}_{\perp})\transpose\mat{X} \\
&=& \mat{V}\boldSigma(\boldSigma^2+\lambda\mat{1}_{|I|})^{-1}\mat{U}\transpose\mat{X}  \\
&=&\mat{V}\boldSigma(\boldSigma^2+\lambda\mat{1}_{|I|})^{-1}(\boldSigma^{-1}\mat{V}\transpose)(\mat{V}\boldSigma\mat{U}\transpose)\mat{X}\\
&=& \mat{V}(\boldSigma^2+\lambda\mat{1}_{|I|})^{-1}\mat{V}\transpose\mat{X}\transpose\mat{X}  \\
&=& (\mat{X}\transpose\mat{X}+\lambda\mat{I}_{|I|})^{-1}\mat{X}\transpose\mat{X}\ \ \ \ \ (=\text{Eq.(\ref{eq:solution1})}).
\end{eqnarray*}
This shows the equivalence of Eqs.(\ref{eq:solution1}) and (\ref{eq:solution2}) when $|U|>|I|$.
\end{proof}

\end{document}